\documentclass[fleqn,usenatbib]{mnras}

\usepackage{newtxtext,newtxmath}
\usepackage{subcaption}
\usepackage[T1]{fontenc}
\usepackage{enumerate}% http://ctan.org/pkg/enumerate
\usepackage{multirow}
\usepackage[labelfont=bf, labelsep=period]{caption}
\DeclareRobustCommand{\VAN}[3]{#2}
\let\VANthebibliography\thebibliography
\def\thebibliography{\DeclareRobustCommand{\VAN}[3]{##3}\VANthebibliography}

\newcommand{\RomanNumeralCaps}[1]
    {\MakeUppercase{\romannumeral #1}}

\usepackage{graphicx}	% Including figure files
\usepackage{amsmath}	% Advanced maths commands
\title[Short title, max. 45 characters]{Cosmography from well-localized Fast Radio Bursts}
\author[J. A. S. Fortunato et al.]{
Jéferson A. S. Fortunato,$^{1}$\thanks{E-mail: jeferson.fortunato@edu.ufes.br}
Wiliam S. Hipólito-Ricaldi,$^{1,2}$
Marcelo V. dos Santos$^{3, 4, 5, 6}$
\\
$^{1}$PPGCosmo, CCE, Universidade Federal do Espírito Santo (UFES), Av. Fernando Ferrari, 540, CEP 29.075-910, Vitória, ES, Brazil\\
$^{2}$ Núcleo Cosmo-UFES \& Departamento de Ciências Naturais, CEUNES, Universidade Federal do Espírito Santo (UFES), Rodovia BR 101 Norte, km. 60,\\
CEP 29.932-540, São Mateus, ES, Brazil\\
$^{3}$Fundação Coppetec, Rua Moniz de Aragão, 360 - Bloco 1 - Cidade Universitária da Universidade Federal do Rio de Janeiro (UFRJ),\\
CEP 21941-594, Rio de Janeiro, RJ, Brazil\\
$^{4}$ Núcleo de Pesquisa Aplicada à Meios Geológicos-Geotécnicos, Av. Athos da Silveira Ramos, 149 - Bloco I, 203 - Cidade Universitária \\da Universidade Federal do Rio de Janeiro (UFRJ), Rio de Janeiro - RJ, 21941-909\\
$^{5}$ Instituto de F\'{i}sica, Universidade de S\~ao Paulo (USP), R. do Mat\~ao, 1371 - Butant\~a, 05508-09 - S\~ao Paulo, SP, Brazil\\
$^{6}$ Unidade Acad\^emica de F\'{i}sica, Universidade Federal de Campina Grande (UFCG), R. Apr\'{i}gio Veloso,  Bodocong\'o, 58429-900 - Campina Grande, PB, Brazil
}

\date{Accepted XXX. Received YYY; in original form ZZZ}

\pubyear{2015}

\begin{document}
\label{firstpage}
\pagerange{\pageref{firstpage}--\pageref{lastpage}}
\maketitle

% Abstract of the paper
\begin{abstract}
Fast Radio Bursts (FRBs) are millisecond-duration pulses occurring at cosmological distances that have emerged as prominent cosmological probes due to their dispersion measure (DM) evolution with redshift. In this work, we use cosmography, a model-independent approach to describe the evolution of the universe, to introduce the cosmographic expansion of the DM-z relation. By fitting two different models for the intergalactic medium and host contributions to a sample of 23 well-localized FRBs, we estimate the kinematic parameters $q_0=-0.59${\raisebox{0.5ex}{\tiny$\substack{+0.20 \\ -0.17}$}}, $j_0=1.08${\raisebox{0.5ex}{\tiny$\substack{+0.62 \\ -0.56}$}}, $s_0=-2.1\pm7.0$, and $H_0=69.4\pm4.7$ achieving a precision of $6\%$ and $7\%$ for the Hubble constant depending on the models used for contributions. Furthermore, we demonstrate that this approach can be used as an alternative and complementary cosmological-model independent method to revisit the long-standing "Missing Baryons" problem in astrophysics by estimating that $82\%$ of the baryonic content of the universe resides in the intergalactic medium, within $7\%$ and $8\%$ precision,  according to the contribution models considered here. Our findings highlight the potential of FRBs as a valuable tool in cosmological research and underscore the importance of ongoing efforts to improve our understanding of these enigmatic events. 
\end{abstract}

% Select between one and six entries from the list of approved keywords.
% Don't make up new ones.

%%%%%%%%%%%%%%%%%%%%%%%%%%%%%%%%%%%%%%%%%%%%%%%%%%

%%%%%%%%%%%%%%%%% BODY OF PAPER %%%%%%%%%%%%%%%%%%

\section{Introduction}
\par Fast Radio Bursts (FRBs) are a class of extremely bright and short-duration transients that occur in the radio spectrum and last for only a few milliseconds. They exhibit large dispersion measures ($\rm DMs$), which are a measure of the electron column density along the sightline. The observed $\rm DMs$ significantly exceed the contribution expected from the Milky Way, providing strong evidence for their extragalactic origin \citep[][]{lorimer2007bright, petroff2019fast}. Since the first FRB discovery in $2007$ by Duncan Lorimer and his team, hundreds of bursts have been reported, and some of these FRBs are known to be repeaters \citep[][]{zhou2022fast}. Among these bursts, $24$ have been precisely localised, and their host galaxy and redshift have been determined. While some FRB events have been linked to magnetars \citep[][]{bochenek2020fast}, numerous progenitor models have been proposed in the literature \citep[][]{zhang2020physical, bhandari2020host}. The cosmological origin of these FRBs has made them a prominent observable in the study of cosmology. 

The so-called "Hubble tension" is a discrepancy between two different methods of estimating the expansion rate of the Universe, known as the Hubble constant ($H_0$). One way of estimating the Hubble constant involves observations of the cosmic microwave background (CMB) radiation, which is a relic of the early universe. The Planck Collaboration \citep[][]{aghanim2020planck} utilised this technique and obtained a value of $H_0=67.4\pm0.5$$\rm \ km \ s^{-1} \ Mpc^{-1}$. Another approach is to measure the distances and parallax using cepheid stars and type Ia supernovae in the local universe to infer $H_0$ directly. This method was employed by \citet{riess2022comprehensive} who found a significantly higher value of $H_0=73.04\pm1.04$$\rm \ km \ s^{-1} \ Mpc^{-1}$, differing by $5\sigma$ from the previous method. In this context, it is important to identify alternative observables that can verify the tension. Fast Radio Bursts have already been used as a tool to infer $H_0$ \citep[][]{macquart2020census, wu20228, zhao2022first, hagstotz2022new}. However, additional data is required to increase the accuracy of the measurement and determine which value is preferred. 

Another cosmological issue that can be addressed by the use of FRBs is the "missing baryons problem". According to measurements obtained by the Planck Collaboration \citep[][]{aghanim2020planck}, the CMB radiation suggests that the vast majority of the Universe, $\sim 95\%$, is comprised of dark energy and dark matter, with only a small fraction, approximately $5\%$, of baryonic matter. However,  in the low-redshift universe has been noted a baryon deficit \citep[][]{fukugita1998cosmic}. This deficit could arise from a not complete understanding of the baryons distribution in the universe and thus, it is important to study the baryon budget. Through a series of observations, \citet{shull2012baryon} suggested that, at low-redshift, roughly $18\%$ of the baryon budget can be accounted for by stars, galaxies, circumgalactic medium (CGM), intracluster medium (ICM), and cold neutral gas, while the remaining $82\%$ exists in a diffuse state within the intergalactic medium (IGM), that is, a fraction of baryons in IGM of $f_{\rm IGM} \approx 0.82$. To cite a few results in literature: using five localised FRBs, \citet{li2020cosmology} estimated a value of $f_{\rm IGM}=0.84$${\raisebox{0.5ex}{\tiny$\substack{+0.16 \\ -0.22}$}}$. The authors in \citep[][]{lemos2022model}, through a model-independent technique and using a sample of 17 FRBs with redshift measurement found $f_{\rm IGM}=0.881\pm 0.012$. In \citep[]{yang2022finding}, the authors obtained $f_{\rm IGM}=0.83\pm 0.06$ using 22 localised FRBs. 

%Finding these missing baryons is important in comprehending the process of galaxy evolution, for instance. 
Aditionally, FRBs have emerged as a potentially powerful tool for investigating a broad range of applications in astrophysics and fundamental physics. These include the detection of the baryon content in the universe \citep[][]{mcquinn2014,deng2014cosmological}, constraints in  equation of state for dark energy \citep[][]{Zhou2014,Gao2014}, trace the magnetic fields in the intergalactic medium \citep[][]{Akahori2016}, test the equivalence principle \citep[][]{Wei2015,Tingay2016,Nusser2016}, and constraint the rest mass of photons \citep[][]{wu2016,shao2017,Lin2023}, among others. Regarding $H_0$ and $f_{IGM}$ constraints, it must be stressed that  most of the cases mentioned above were performed for a particular cosmological model: the $\Lambda$CDM. However, we would get a complementary perspective if any model-independent method were used.

One of the most noticeable model-independent approaches in  cosmology is the cosmography. This method, which has been extensively discussed in previous works \citep[][]{weinberg1972gravitation, visser2005cosmography, aviles2012cosmography}, relies solely on the cosmological principle and employs the FLRW metric. Unlike other approaches that require the use of Friedmann equations derived from General Relativity, the cosmography expands observable quantities such as luminosity distance into power series and establishes a direct relationship between cosmological parameters and the available data. This approach has been extensively utilised in prior research employing a range of cosmological probes including Baryon Acoustic Oscillations, Type Ia supernovae, and Cosmic Chronometers \citep[][]{lazkoz2013bao, riess2022comprehensive, jalilvand2022model}. However, it has not yet been applied to Fast Radio Bursts.

Fast Radio Bursts  dispersion measure (DM) has been suggested as a possible and complementary tool to other established techniques in cosmology. It is interesting to explore the application of cosmographic expansion to the DM and examine the insights that FRBs can offer regarding the cosmographic approach. The present paper applies cosmography to the  FRBs dispersion measure-redshift ($\rm DM - z$) relationship to determine some cosmographic parameters (deccelatarion parameter $q_0$, jerk parameter $j_0$ and snap parameter $s_0$), including the Hubble constant using the most recent set of well-localised FRBs. Furthermore, we estimate the fraction of baryons present in the intergalactic medium (IGM).  The paper is structured as follows: In Section \ref{2}, we discuss the fundamental properties of FRBs. In Section \ref{3}, we introduce the cosmographic equations. Section \ref{4} outlines  methods for the likelihood computation. We present our data and results in Section \ref{5}, and finally, we discuss the results in Section \ref{6}.

\section{Properties of FRBs}\label{2}

During its path to earth, a FRB pulse is dispersed by the intergalactic medium. The amount of dispersion is given by the time delay of  different radio frequencies that compose the signal observed:
\begin{equation}
    \Delta t\propto \left(\nu_{\mathrm{lo}}^{-2} - \nu_{\mathrm{hi}}^{-2}\right) \mathrm{DM},
\end{equation}
\noindent where $\nu_{\mathrm{lo}}$ and $\nu_{\mathrm{hi}}$ represents the low and high frequencies respectively. The dispersion measure $\mathrm{DM}$ is related to the column density of free electrons $n_e$ along the FRB line of sight $l$ weighted by redshift as
\begin{equation}
    \mathrm{DM} = \int \frac{n_e}{(1+z)}dl.
\end{equation}

The observed $\rm DM_{obs}$ includes two main contributions from the intergalactic and extra-galactic mediums, $\mathrm{DM}_\mathrm{local}$ and $\mathrm{DM}_{\mathrm{EG}}$:
\begin{equation}\label{dmobs}
    \rm DM_{obs} = \mathrm{DM}_\mathrm{local} + \mathrm{DM}_\mathrm{EG}(z),
\end{equation}
\noindent being 
\begin{equation}
    \mathrm{DM}_{\mathrm{local}}=\mathrm{DM}_{\mathrm{ISM}}+\mathrm{DM}_{\mathrm{halo}},
\end{equation}
\noindent and
\begin{equation}
    \mathrm{DM}_{\mathrm{EG}} = \mathrm{DM}_{\mathrm{IGM}}+\frac{\mathrm{DM_{\mathrm{host}}}}{(1+z)},
\end{equation}
\noindent where $\mathrm{DM}_{\mathrm{ISM}}$ corresponds to the contributions from the Milky Way interstellar medium, usually calculated through galactic electron distribution models such as $\mathrm{NE2001}$ \citep[][]{cordes2002ne2001} and $\mathrm{YMW16}$ \citep[][]{yao2017new} then subtracted from the observed $\mathrm{DM}$.  Here we used NE2001 approach because in recent works it has been  reported that YMW16 model may overestimate $\mathrm{DM}_{\mathrm{ISM}}$ at low Galactic latitudes \citep[][]{koch2021}. On the other hand,  $\mathrm{DM}_{\mathrm{halo}}$  is related to Milk Way galactic halo which has been estimated  in the range of $ 50 <  \mathrm{DM}_{\mathrm{halo}}<100~\mathrm{pc}\,\mathrm{cm}^{-3}$ \citep[][]{prochaska2019probing}. However, to be conservative we assume $\mathrm{DM}_{\mathrm{halo}}  = 50~\mathrm{pc}\,\mathrm{cm}^{-3} $ as for example in \citep[][]{macquart2020census}. $\mathrm{DM}_{\mathrm{IGM}}$ is the contribution from the intergalactic medium ($\mathrm{IGM}$) which has cosmological dependence and $\mathrm{DM}_{\mathrm{host}}$ is the host galaxy component corrected with $(1+z)^{-1}$ to account for cosmological expansion for a FRB source at redshift $z$.

%\subsection{Dispersion measure from the intergalactic medium}
%\label{dispc} % used for referring to this section from elsewhere

The dominant contribution to the observed dispersion measure of a FRB signal is due to the intergalactic medium ($\mathrm{IGM}$). In a previous study, \citet{mcquinn2013locating} reported that this component is responsible for an expressive scatter around the mean $\mathrm{DM}_{\mathrm{IGM}}$, $100$ -- $400 \mathrm{pc}\,\mathrm{cm}^{-3}$ at $z = 0.5$ -- $1$, respectively. Following \citet{deng2014cosmological}, the average $\mathrm{DM}_\mathrm{IGM}$ is given by
\begin{equation}\label{DM_igm}
{\rm \langle DM_{IGM} \rangle } = {\rm A\Omega_b} H_0\int^z_0\frac{(1+z)f_{\rm IGM}(z)f_{\rm e}(z)}{E(z^\prime)}dz^\prime,
\end{equation}
\noindent where $E(z)=H(z)/H_{0}$, $f_{\rm e} = Y_{\rm
H}X_{\rm e,H}(z)+\frac{1}{2}Y_{\rm He}X_{\rm e,He}(z)$
and $A=3c/8\pi Gm_{\rm p}$. The cosmic baryon density, the mass of proton and the fraction of baryon mass in the $\rm IGM$ are represented by $\rm \Omega_b$, $m_{\rm p}$ and $f_{\rm IGM}$, respectively. In this work we also consider an $\rm IGM$ with a hydrogen mass fraction $Y_{\rm
H}=0.75$ and a helium mass fraction $Y_{\rm He}=0.25$. Given the fact that hydrogen and helium are completely ionized at $z<3$, the ionization fractions of each species are $X_{\rm e,H}=X_{\rm e,He}=1$. For this analysis, first, we keep a constant value for the fraction of baryon mass, $f_{\rm IGM} = 0.82$ \citep[][]{shull2012baryon} but after we leave it as a free parameter.

According to equation (\ref{dmobs}), $\mathrm{DM}_{\mathrm{IGM}}$ is estimated by $\mathrm{DM}_{\rm IGM} = \rm DM_{\rm obs} - \rm DM_{\rm local} - \rm DM_{\rm host}(1+z)^{-1}$ with uncertainty given as 
\begin{equation}\label{error}
    \sigma_{\rm IGM}(z)=\sqrt{\sigma_{\rm obs}(z)^2+\sigma^2_{\rm local}+\left(\frac{\sigma_{\rm host}(z)}{1+z}\right)^2},
\end{equation}
where $\sigma_{\rm obs}$ and $\sigma_{\rm host}$ are the errors related to $\rm DM_{\rm obs}$ and $\rm DM_{\rm host}$, respectively. Whereas $\sigma_{\rm local}$ is the sum of $\rm DM_{\rm ISM}$ and $\rm DM_{\rm halo}$ uncertainties,  we follow \citet{hagstotz2022new} taking $\sigma_{\rm local} \approx 30~\rm{pc/cm}^{-3}$.

The scatter around the averaged quantity ${\rm \langle DM_{IGM} \rangle }$ is due to inhomogeneities in  the column density of free electrons along the FRB line of sight.  The distribution of ${\rm DM_{IGM}}$ is greatly influenced by the way how baryons are distributed around galactic halos as shown by cosmological simulations. The number of collapsed structures that a given line of sight intersects, it determines the extent of variation around ${\rm \langle DM_{IGM} \rangle }$ thus, more compact halos leads to a skewed probability distribution associated with $\rm {\langle DM_{IGM} \rangle }$, while a more diffuse gas around the halos results in a more Gaussian-like probability distribution function \citep[][] {macquart2015fast, macquart2020census, bhandari2021}. 
In the literature, both approaches have been studied. For example in \citep[][]{prochaska2019frbs, yang2022finding,wu2020new}, non-gaussian approach was used, while in \citep[][]{jaroszynski2019fast, hagstotz2022new, zhang2023cosmology}, considered the gaussian point of view. In this work, we use these two  approaches to model the IGM component looking forward to detect any possible different impact on final conclusions.

The first approach we use is more conservative as it assumes a gaussian distribution around the mean, given by equation (\ref{DM_igm}), with standard deviation interpolated in the range $\sigma_{\rm IGM}(z=0)\approx 10 ~\mathrm{pc}\,\mathrm{cm}^{-3}$ and $\sigma_{\rm IGM}(z=1)\approx 400 ~\mathrm{pc}\,\mathrm{cm}^{-3}$. This method was used, for instance, in \citet{hagstotz2022new}. In the second approach, we  follow \citet{macquart2020census} and assume a quasi-gaussian distribution for the IGM contribution:
\begin{equation}\label{PIGM}
P_{\rm IGM}(\Delta) = A \Delta^{-\beta} \exp  \left[-\frac{(\Delta^{-\alpha}-C_0)}{2\alpha^2\sigma^2_{\rm DM}} \right],  ~~~ \Delta > 0, 
\end{equation}
being $\Delta\equiv \rm DM_{IGM}/\langle DM_{IGM}\rangle$. This approach, which has shown excellent agreement with the observed distributions of $\rm DM_{IGM}$ in both semi-analytic models and hydrodynamic simulations, is based on the fact that when the variance $\rm \sigma_{DM}$ is large, it captures the skewness due to the different sightlines that cross a few large structures increasing the $\rm DM_{IGM}$ value. Conversely, in the limit of small $\rm \sigma_{DM}$, the distribution of Eq.~(\ref{PIGM})  becomes Gaussian. The parameters $\alpha$ and $\beta$ are related to the inner density profile of gas inside galactic halos. We use the values from \citet{macquart2020census}, $\alpha = 3$ and $\beta = 3$. The remaining two parameters $A$ and $C_0$ are fitted when $\Delta = 1$. The standard deviation with redshift of $\rm DM_{\rm IGM}$ can be estimated by
\begin{equation}\label{stand}
    \sigma_{\rm DM} = Fz^{-0.5},
\end{equation}
where $F$ quantifies how strong is the baryon feedback, that is, how diffuse is the gas around the halo. Following \citet{macquart2020census}, we assume $F=0.32$. 
%\section{The properties of localized FRBs}\label{sec2}

%\subsection{Dispersion measure from the host galaxy}\label{hostgal}

Although the dispersion measure of the   host environment $\rm DM_{\rm host}$  is a crucial feature to determine the source, it still has few theoretical motivations. In order to estimate this component, informations about the host galaxy type, electron distribution, position of FRB signal within the galaxy, and the viewing angle are required. However, 
 all of these information are still uncertain and thus we focus on the probability distribution of $\rm DM_{\rm host}$. Here we consider two cases for modeling the $\rm DM_{\rm host}$: the first one, following \citet{hagstotz2022new}, it is based on the stochastic contribution:
 \begin{equation}
    P(\mathrm{DM}_\mathrm{host}) = \mathcal{N} \big( \langle \mathrm{DM}_\mathrm{host} \rangle, \sigma_\mathrm{host}^2 \big) \; ,
\end{equation}
being $\mathcal{N}$ a normal distribution. For this approach, we assume galactic halos similar to the Milky Way, then for the mean value we have $\langle \mathrm{DM}_\mathrm{host} \rangle = 100 (1+z_\mathrm{host})^{-1} \, \mathrm{pc}\,\mathrm{cm}^{-3}$ and for the variance, $\sigma_\mathrm{host} = 50 (1+z_\mathrm{host})^{-1} \, \mathrm{pc}\,\mathrm{cm}^{-3}$. 

The second case, we follow \citet{macquart2020census} and consider a log-normal distribution, as it has a long asymmetric tail allowing for large $\rm DM_{\rm host}$ values:
\begin{equation}\label{Phost}
P({\rm DM_{host}}) = \frac{1}{{\rm DM_{host}} \sigma_{\rm host} \sqrt{2\pi}} {{\exp}}\left(-\frac{{\rm ln} {\rm DM_{host}}-\mu}{2\sigma_{\rm host}^2}\right),
\end{equation}
where $e^{\mu}$ and $e^{2\mu+\sigma_{\rm host}^2}(e^{\sigma_{\rm host}^2}-1)$ are the mean and variance of the distribution, respectively. %\textbf{The free arameters $\mu$ and $\sigma_{\rm host}$ were fitted through the state-of-the-art IllustrisTNG simulations \citep{zhang2020b}.}

%\subsection{Milky Way contribution}

%The Milky Way interstellar medium component $\rm DM_{\rm ISM}$ can be obtained by electron density models and further subtracted this contribution for each FRB observed $\rm DM$. In addition, one has to account for the Milky Way's circumgalactic medium ($\rm CGM$), $\rm DM_{\rm halo}$. In this work, we consider a constant value of $\rm DM_{\rm halo} = 50 ~ \mathrm{pc}\,\mathrm{cm}^{-3}$ with negligible dispersion.

%[\centering label 1]

\section{Cosmography with fast radio bursts}\label{3}

Cosmography, or cosmic kinematics, plays an important role when studying cosmic expansion in a model-independent way, i.e.,
without dependence on any specific model for the underlying cosmic evolution. This approach is based  on the cosmological principle, which postulates the homogeneity and isotropy of the Universe on large scales. In order to describe the kinematics of the cosmic expansion, one needs to use the Hubble parameter:
\begin{equation}\label{hubblep}
    H(t)=\frac{1}{a}\frac{da}{dt},
\end{equation}
and additionally the functions \citep[][]{visser2005cosmography},
\begin{equation}\label{dp}
    q(t)=-\frac{1}{a}\frac{d^2a}{dt^2}\left(\frac{1}{a}\frac{da}{dt}\right)^{-2},
\end{equation}

\begin{equation}
    j(t)=\frac{1}{a}\frac{d^3a}{dt^3}\left(\frac{1}{a}\frac{da}{dt}\right)^{-3},
\end{equation}
and
\begin{equation}
    s(t)=-\frac{1}{a}\frac{d^4a}{dt^4}\left(\frac{1}{a}\frac{da}{dt}\right)^{-4}.
\end{equation}
The cosmological expansion rate is characterized by $H(t)$, while the deceleration parameter, $q(t)$, represents the acceleration or deceleration of  the universe's expansion. The jerk parameter $j(t)$ can be used to estimate if there was a transition period  in which the universe modified its expansion by changing the rate of the expansion acceleration. Additionally, the snap parameter, $s(t)$, is important to discriminate between a cosmological model that allows an evolving dark energy term or one with a cosmological constant. Although there exists also other parameters that includes higher-order time derivatives of the scale factor, we focus only in $H(t)$, $q(t)$, $j(t)$ and $s(t)$. 

By using the parameters defined above, scale factor can be expanded around the present time, $t_0$ as
\begin{eqnarray} \label{seriea}
    \frac{a(t)}{a_0} &=& 1+H_0 (t-t_0) -\frac{1}{2}q_0H^2_0 (t-t_0)^2\\\nonumber\\
    &+&\frac{1}{3!}j_0 H^3_0 (t-t_0)^3 + \frac{1}{4!}s_0H_0^4(t-t_0)^4+O\left[(t-t_0)^5\right].\nonumber
\end{eqnarray}
From now on, we set $a_ 0=1$ and $H_0$, $q_ 0$, $j_ 0$ and $s_0$ stand for the quantities evaluated in current time $t_ 0$. Eq. (\ref{seriea}) helps us to find the cosmographic series for $E(z)$-function, 
with which it is possible to obtain the expansion for luminosity distance, angular distance, redshift drift (see for example \citet{lobo2020,heinesen2021,Pourojaghi2022,rocha2023}) or in our case, the $\mathrm{DM}_{\mathrm{IGM}}(z)$ given by eq.(\ref{DM_igm}). To do this we use the relation:
\begin{equation}
    \frac{dH}{dz}=\left(\frac{1+q}{1+z}\right)H\,,
\end{equation}
and obtain the $\rm DM - z$ relation in terms of cosmographic parameters,
\begin{eqnarray}\label{cosmog}
     \mathrm{DM}_{\mathrm{IGM}}(z)&=&\frac{3\Omega_b H_0^2}{8 \pi Gm_p}f_e f_{\rm IGM}\biggl\{ \frac{cz}{H_0}\biggl[ 1-\frac{q_0 z}{2} \\\nonumber
     &+&\frac{1}{6}\left(4+6q_0+3q_0^3-j_0\right)z^2\\\nonumber
     &-&\frac{1}{24}\left(18q_0+42q_0^2+14q_0j_0-9q_0^3-s_0-18j_0\right)z^3\\\nonumber
     &+&O(z^4)\biggl] \biggl\}.
\end{eqnarray}
The Taylor series, as shown in eq.(\ref{cosmog}), exhibits convergence issues for high-redshifts ($z > 1$). This concern prompted \cite{cattoen2007} to address the convergence problems and propose a new parametrization, denoted as $y=z\left(1+z\right)^{-1}$. Subsequently, alternative parametrizations utilising Padé expansions \citep{aviles2014}, Chebyshev polynomials \citep{capozziello2018}, and logarithmic polynomials \citep{bargiacchi2021} have been proposed. A comprehensive comparison of these methodologies was conducted in \cite{Hu2022}. However, given that the localized FRBs data we utilise falls within the range $0.0039 < z < 0.66$, the $z$-parametrization is suitable for our analysis.

It is worth emphasizing that the parameters $H_0$, $q_0$, $j_0$, and $s_0$ are solely defined within the cosmographic framework and do not inherently relate to any specific cosmological content. To establish a connection between these parameters and the characteristics of a particular cosmological model, the Einstein equations, specifically the Friedmann equations, must be additionally considered. However, in this study, our objective does not involve relating the cosmographic parameters to any specific cosmological model. Instead, we focus on directly measuring the expansion through kinematic quantities and mapping its temporal evolution. Consequently, our set of free parameters is denoted as $\theta=\{H_0, \Omega_b h^2, q_0, j_0, s_0, f_{\rm IGM}\}$.

\section{Method} \label{4}
The main purpose of this study is to verify how cosmography, or cosmic kinematics, might be helpful when constraining cosmic expansion by using Fast Radio Bursts along with its redshift measurements. In this context, first, we define two models based on the assumptions taken for the distributions of $\rm DM_{\rm IGM}$ and $\rm DM_{host}$, in the first one we follow \citet{hagstotz2022new}, and in the second one we follow \citet{macquart2020census}:

\begin{enumerate}[(I)]%[label=(\roman*)]
        \item For the first model, we consider gaussian distributions for both $\rm DM_{\rm IGM}$ and $\rm DM_{host}$ and then, every observed dispersion measure $\rm DM_i$ at a redshift $z_i$ will be related to a gaussian individual likelihood,
        \begin{equation}
            \mathcal{L}(\rm{DM}_i, z_i) = \frac{1}{\sqrt{2 \pi \sigma_i^2}} \exp \left[ \frac{\bigl(\rm{DM}_i - \rm{DM}^{\rm{theo}}(z_i) \bigr)^2} {2 \sigma_i^2} \right] \,,
        \end{equation}
        where $\rm{DM}^{\rm theo}(z_i)$ is the theoretical contribution as stated in section \ref{2}, 
        \begin{eqnarray}\nonumber\label{dmteo}
        \rm{DM}^{\rm{theo}}(z_i) &=&   \rm DM_{obs} - \rm DM_{\rm ISM}-\rm{DM}_{halo}\\ 
        &=&  \rm{DM}_{\rm{IGM}} (z_i) +    \rm{DM}_{\rm{host}} (z_i) \, .
        \end{eqnarray}
        
        The effect of measurement errors on $\rm DM_{\rm IGM}$ is minimal, and as a result, the overall variation is determined by the individual uncertainties which encompass the spread from the intergalactic medium contribution, the Milky Way electron distribution model, and the host galaxy: 
        \begin{equation}
            \sigma^2(z_i) =  \sigma_{\rm{MW}}^2 +       \sigma_{\rm{host}}^2(z_i) +       \sigma_{\rm{IGM}}^2(z_i) \, .
        \end{equation}
        Given that all events are independent, the combined likelihood of the sample is simply the multiplication of the separate likelihoods:
        \begin{equation}
        \label{eq:likelihood_tot}
        \mathcal{L}_{\rm{tot}} = \prod_i  \mathcal{L}_i \: ,
        \end{equation}
        and the computation of the product is executed for every FRB listed in Table \ref{data}.
        
        \item For the second model, we consider that the distribution of $\rm DM_{\rm IGM}$ is quasi-gaussian distributed according to eq.~(\ref{PIGM}), and for the $\rm DM_{host}$ we assume the lognormal distribution given by eq.~(\ref{Phost}). The total probability density function of a FRB being detected at a redshift $z_i$ with $\rm{DM}^{\rm{theo}}(z_i)$ given by eq.~( \ref{dmteo}) is determined by the following relation:
        \begin{eqnarray}\nonumber
        P_i(\rm{DM}^{\rm{theo}}(z_i))  &=&
        \int\limits_0^{\rm{DM}^{\rm{theo}}} \;
        P_{\rm host}({\rm DM}_{\rm host}) \;\\
        &\times& P_{\rm IGM}({\rm DM_{\rm IGM}}) \;       d{\rm DM}_{\rm host}  \;\; ,
        \label{eqn:prob}
        \end{eqnarray}
        with $P_{\rm host}({\rm DM}_{\rm host})$ and $P_{\rm IGM}({\rm DM_{\rm IGM}})$ being the probability density functions for $\rm DM_{host}$ and $\rm DM_{\rm IGM}$ as described in section \ref{3}. Finally, we calculate the joint likelihood function by combining the probability density functions of each FRB through the product:
        \begin{equation}
        \mathcal{L}_{\rm{tot}} = \prod_i P_i(\rm{DM}^{\rm{theo}}(z_i)) \: .
        \end{equation}       
\end{enumerate}
In this study, we used the Nested Sampling algorithm -- via the publicly available Python package \verb|Polychord| \citep[][]{handley2015polychord}, which is a Monte Carlo (MC) technique, to place constraints on $H_0$, $\Omega_bh^2$, $q_0$, $j_0$, $s_0$ and $f_{\rm IGM}$ parameters. We have implemented these methods to a sample of $23$ localised FRBs using the two models described above (\RomanNumeralCaps{1} and \RomanNumeralCaps{2}) in order to estimate the set of best-fit parameters for each model. Nested sampling is a powerful method for Bayesian parameter estimation because it has several advantages over other methods, such as Markov Chain Monte Carlo (MCMC). It enables us to extensively explore the parameter space and to accurately determine the probability distributions of the parameters of interest. It also provides a way to compute the evidence of a model, i.e., the probability of the data given a model, which is important for model comparison and selection. Furthermore, certain calculations presented in this paper were made feasible by modifying the publicly available Python code, \verb|FRB| \citep[][]{prochaska2019frbs}.

\begin{figure}
 \includegraphics[width=\columnwidth]{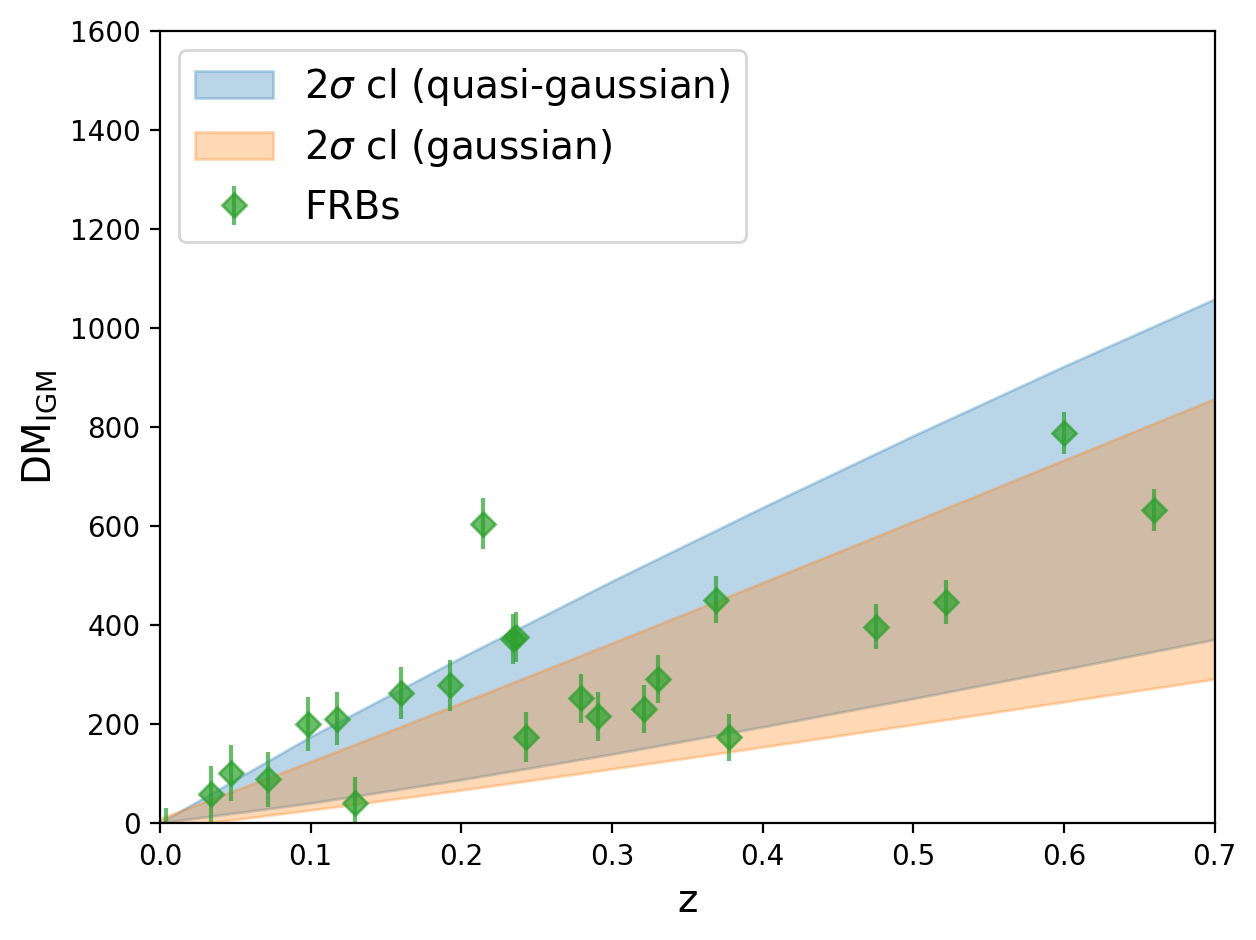}
 \caption{The $2\sigma$ confidence level region for the two distribution models considered in this work  plotted as shaded areas. The blueish represents the quasi-gaussian distribution as described by Eq.~(\ref{PIGM}) and the orangish represents the gaussian one. The well-localised FRBs are plotted as scatters.}
 \label{scatters}
\end{figure}

%%%%%%%%%%%%%%%%%%%%%%%%%%%%%%%
 \section{Data and results}\label{5}
%%%%%%%%%%%%%%%%%%%%%%%%%%%%%
\subsection{Localized Fast Radio Bursts}

\begin{table}
	\small
	\begin{center}
		\caption{ Properties of localised FRBs.}
		\begin{tabular}{ccccc}
			\hline
			Name & Redshift & DM$_{\rm obs}$   & Reference \\
			&  & $(\rm pc \ cm^{-3})$  & &  \\
			\hline
			FRB 121102 &    0.19273 &$557$	    & \cite{spitler2016repeating}\\
			FRB 180301 &    0.3304  &$534$           & \cite{bhandari2022characterizing}\\
			FRB 180916 &	0.0337  &$348.8$	      & \cite{marcote2020repeating}\\
			FRB 180924 &    0.3214  &$361.42$        & \cite{bannister2019single}\\
			FRB 181030 &    0.0039  &$103.5$      & \cite{bhardwaj2021local}\\
			FRB 181112 &    0.4755	&$ 589.27$      & \cite{prochaska2019low}\\
			FRB 190102 &    0.291 	&$ 363.6$	       & \cite{bhandari2020host}	\\
			FRB 190523 &    0.66    &$ 760.8$        & \cite{ravi2019fast}\\
			FRB 190608 &    0.1178  &$ 338.7$	       & \cite{chittidi2021dissecting}	\\
			FRB 190611 & 	0.378   &$ 321.4$	     & \cite{day2020high}  \\
			FRB 190614 &    0.6     &$959.2$            & \cite{law2020distant}\\
			FRB 190711 &    0.522 	&$ 593.1$     & \cite{heintz2020host}\\
			FRB 190714 &    0.2365	&$ 504.13$         & \cite{heintz2020host}\\
			FRB 191001 &	0.234   &$ 506.9$        & \cite{heintz2020host}\\
			FRB 191228 &    0.2432  &$ 297.5$         & \cite{bhandari2022characterizing}\\
			FRB 200430 &    0.16	&$ 380.25$      & \cite{heintz2020host}\\
			FRB 200906 &    0.3688  &$ 577.8$          & \cite{bhandari2022characterizing}\\
			FRB 201124 & 	0.098   &$ 413.52$      & \cite{fong2021chronicling}\\
            FRB 210117 & 0.2145 & $730.0$& \cite{james2022measurement} \\
            FRB 210320 & 0.2797 & $384.8$& \cite{james2022measurement} \\
		FRB 210807 & 0.12927 & $251.9$ & \cite{james2022measurement}\\
		FRB 211127 & 0.0469 & $234.83$ & \cite{james2022measurement} \\
		FRB 211212 & 0.0715 & $206.0$ & \cite{james2022measurement} \\
			\hline
			% \toprule
			% \toprule
		\end{tabular}
		\label{data}
		\vspace{0.5cm}
	\end{center}
\end{table}

To date, numerous Fast Radio Bursts (FRBs) have been documented by various collaborations, with a total count surpassing $1000$. However, a relatively small fraction of these FRBs, specifically $24$, have measurements of their redshift available. In our analysis, we focus on a subset of these FRBs, compiled by \citet{yang2022finding}, precisely the $23$ instances listed in Table \ref{data}, which includes both their redshift and observed dispersion measure (DM) values. It is worth mentioning that some of these FRBs exhibit a repeating behavior. This characteristic enables the possibility of detecting their source through interferometry techniques \citep[][]{chatterjee2017direct}. The first documented repeating FRB, named FRB 121102, was originally identified by observations made using the Arecibo radio telescope \citep[][]{spitler2016repeating}. Since then, numerous subsequent pulses have been detected from this particular source, with a previous study noting a period of activity lasting for approximately $157$ days \citep[][]{rajwade2020possible}. Notably, among the localized FRBs, a subset of 7 FRBs has been reported to exhibit repeating behavior. In our analysis we chose to exclude the nearest point: FRB200110E as it carries little cosmological information. This is the closest extra-galactic FRB detected so far, located in the M81 galaxy, only $3.6~\rm Mpc$ distant from Earth \citep[][]{bhardwaj2021nearby, kirsten2021repeating}. 

To start the analysis, first, we have to estimate $\rm DM_{\rm IGM}$ for the two models considered here. The estimated $\rm DM_{\rm IGM}$ of the localised FRBs are displayed in Fig.~\ref{scatters} as scatters along with their error bars calculated using the eq.~(\ref{error}). The greenish-shaded area refers to the $95\%$ confidence region for the scatter around the mean intergalactic medium $\rm DM$ component considering the dispersion from eq.(\ref{PIGM}), and the yellowish-shaded  represents the gaussian distribution as described in section \ref{2}. In both cases, FRB 210117 is off the $95\%$ region, which is possibly caused by its local environment contribution to the dispersion measure \citep[][]{yang2022finding}. We chose to keep this data point in our calculations despite this outlier feature because  we did not detect any relevant  difference with our final results by retiring this point.

%The black dashed line represents the averaged $\rm DM_{\rm IGM}$ value calculated from eq.~( \ref{cosmog}) \textbf{with best-fit values for $H_0$, $q_0$ and $j_0$}, and  the dashed-dotted line refers to the averaged relation from the $\Lambda$CDM model \textbf{ with best-fit parameters found by SHOES (\textcolor{red}{referenecias?})}

To ensure the independent and accurate determination of the parameters in Eq.~(\ref{cosmog}), it is essential to constrain $H_0$, $\Omega_b h^2$, and $f_{\rm IGM}$ as independently as possible. Here we have explored two distinct scenarios: a) In the first case, we have imposed a narrow prior on $\Omega_b h^2$, consistent with observations from primordial nucleosynthesis (BBN) and the cosmic microwave background (CMB). We have fixed the value of $f_{\rm IGM}$ and treated $H_0$ as a free parameter. The primary objective of this analysis is to derive information about $H_0$, $q_0$, $j_0$, and $s_0$. b) In the second case, we have imposed a narrow prior on both $\Omega_b h^2$ and $H_0$, while allowing $f_{\rm IGM}$ to be a free parameter. Here, our focus is on obtaining insights into $f_{\rm IGM}$, $q_0$, $j_0$, and $s_0$. In both cases a) and b) we use the the following flat priors: $\left[-0.95, -0.3 \right]$ for $q_0$, $\left[0, 2\right]$ for $j_0$ and  $\left[-5, -10\right]$ for $s_0$.

%\textcolor{red}{To provide a sound justification for our choice of priors in this study, it is necessary to present the $\Lambda \rm CDM$ cosmographic parameters}:

%\begin{eqnarray}
 %   q_0&=&-1+\frac{3}{3}\Omega_m;\\
  %  j_0&=&1; \\
   % s_0&=&1-\frac{9}{2}\Omega_m,
%\end{eqnarray}

%\textcolor{red}{\noindent where $\Omega_m$ represents the matter density evaluated at $z=0$. In line with previous works such as \citet{lizardo2021cosmography}, \citet{dunsby2016theory}, and \citet{bolotin2018applied}, by choosing $\Omega_m=0.315$, one has $q_0=-0.52$, $j_0=1$ and $s_0=-0.39$. Then we adopt their approach to establish the framework for our analysis considering the priors presented in table \ref{priors}. With this foundation, we can now delve into the specific results obtained for each case individually.}

%\begin{table}
%\centering
%\caption{Priors for the cosmographic parameters used in our calculations.}
%\begin{tabular}{cc}
%\hline
%\textbf{Parameter} & \textbf{Prior} \\
%\hline
%$q_0$ & Flat in $\left[-0.95, -0.3 \right]$ \\
%$j_0$ & Flat in $\left[0, 2\right]$ \\
%$s_0$ & Flati in $\left[-5, -10\right]$ \\
%\hline
%\end{tabular}
%\label{priors}
%\end{table}

\begin{figure*}
 \includegraphics[width=15cm]{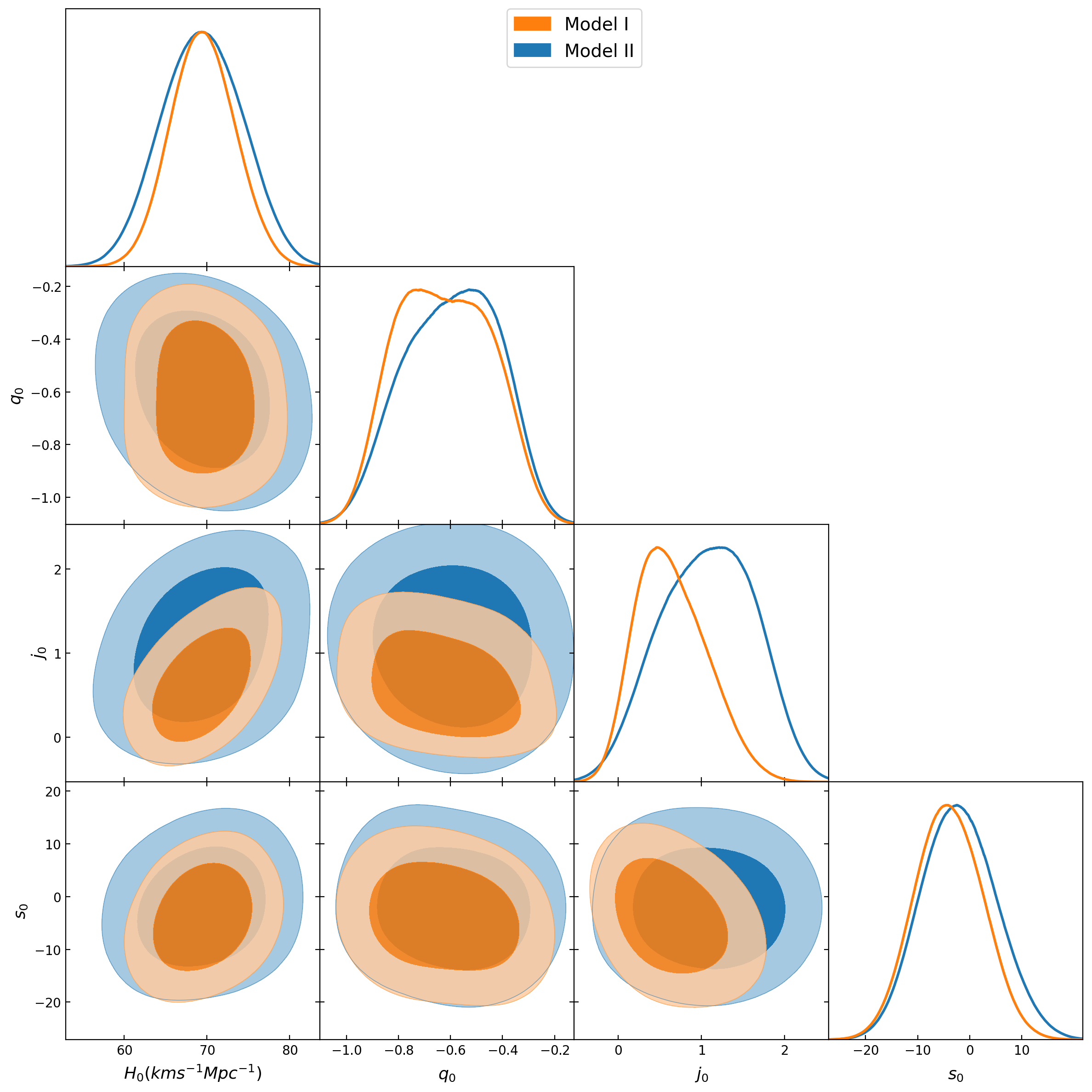}
 \caption{Comparison between the constraints for the two models considered in this work using the 23 well-localised FRBs sample when fixing $f_{\rm IGM}$ and considering a gaussian prior on $\Omega_b h^2$.}
 \label{gcont1}
\end{figure*}

In case a), two types of priors were considered for $\Omega_b h^2$, taking into account consistency with Big Bang nucleosynthesis (BBN) and the cosmic microwave background (CMB). The first prior is a flat prior within the interval $\left[0.02186, 0.02284\right]$, which aligns with findings from previous studies \citep[][]{aghanim2020planck, cooke2018one}. The second prior is a Gaussian prior with $\Omega_b h^2 = 0.02235 \pm 0.00049$, in agreement with the value reported by \citet[][]{cooke2018one}. Additionally, a flat prior on $H_0$ within the range $\left[50, 80\right]$ was applied, and the value of $f_{\rm IGM}$ was fixed at 0.82, as estimated by \citet{shull2012baryon}. The results for this case are presented in Table \ref{results1} and Fig. \ref{gcont1} for both Models I and II.

\begin{center}
\captionof{table}{The constraints for the cosmographic parameters of the $\rm DM - z$ relation when considering different priors over $\Omega_b h^2$. The results for $H_0$ are displayed with $6\%$ and $7\%$ precision.}
\label{results1}
\resizebox{\linewidth}{!}{
\begin{tabular}{c c c c c c}
\hline%\toprule
Prior on $\Omega_b h^2$ & Model & $H_0$ & $q_0$ & $j_0$ & $s_0$\\
\hline%\midrule
\multirow{2}{*}{Gaussian} & \RomanNumeralCaps{1} & $69.5${\raisebox{0.5ex}{\tiny$\substack{+3.6 \\ -4.0}$}} & $-0.63 \pm 0.18$& $0.64${\raisebox{0.5ex}{\tiny$\substack{+0.36 \\ -0.50}$}} & $-4.0\pm6.3$\\
& \RomanNumeralCaps{2} & $69.4\pm 4.7$ & $-0.59${\raisebox{0.5ex}{\tiny$\substack{+0.20 \\ -0.17}$}} & $1.08${\raisebox{0.5ex}{\tiny$\substack{+0.62 \\ -0.56}$}} & $-2.1\pm 7.0$\\
 \hline
\multirow{2}{*}{Flat} & \RomanNumeralCaps{1} & $70.0\pm3.9$ & $-0.61\pm0.19$ & $0.72${\raisebox{0.5ex}{\tiny$\substack{+0.39 \\ -0.51}$}} & $-4.6${\raisebox{0.5ex}{\tiny$\substack{+6.8 \\ -6.0}$}}\\
& \RomanNumeralCaps{2} & $70.0\pm 5.3$ & $-0.60 \pm 0.19$ & $1.10${\raisebox{0.5ex}{\tiny$\substack{+0.66 \\ -0.55}$}} & $-2.4\pm7.1$\\
\hline
%\bottomrule
\end{tabular}}
\end{center}\vspace{1cm}
 
The determined values of $H_0$ can be observed in Table \ref{results1}: $H_0=$ $69.5$${\raisebox{0.5ex}{\tiny$\substack{+3.6 \\ -4.0}$}}~ {\rm  \ km \ s^{-1} \ Mpc^{-1}}$ and $H_0=$ $69.4$${\raisebox{0.5ex}{\tiny$\substack{+5.4 \\ -4.8}$}}~ {\rm  \ km \ s^{-1} \ Mpc^{-1}}$ for the Gaussian $\Omega_b h^2$ prior, and $H_0=70.0\pm3.9$$~{\rm  \ km \ s^{-1} \ Mpc^{-1}}$ and $H_0=70.0\pm5.3$$~{\rm  \ km \ s^{-1} \ Mpc^{-1}}$ for the flat prior case. All four values exhibit good agreement within the $1\sigma$ statistical confidence level, both among themselves and with previous results reported, such as those derived from Supernovas type Ia \citep[][]{riess2022comprehensive}, Cosmic Chronometers \citep[][]{gomez2018h0}, and recent FRB-based investigations within the $\Lambda$CDM framework \citep[][]{wu20228, zhao2022first}. Nevertheless, the large error bars associated with these values limit their informativeness regarding the Hubble tension. In terms of precision, the use of the 23 FRB data points allows for a $H_0$ precision of approximately $\sim 6\%$ for Model I and $\sim 7\%$ for Model II.

The results for the deceleration parameter $q_0$ presented in Table \ref{results1} exhibit agreement within a $1\sigma$ confidence level, indicating an accelerated expansion. In fact, a decelerated phase is rejected at a significance level of approximately $3.2 \sigma$, corresponding to a probability of approximately $99.97\%$. These findings are particularly interesting as they are derived solely from the analysis of FRBs data. However, it is important to note that the error bars for $q_0$ are relatively large. While our results suggest a precision of around $32\%$ for this parameter, it decreases significantly when considered in conjunction with other observational probes (see below). Similar to $q_0$, the jerk parameter $j_0$ also demonstrates agreement within a $1\sigma$ confidence level for all values reported in Table \ref{results1}. Notably, the $\Lambda$CDM value of $j_0=1$ lies within the $1\sigma$ interval for all cases, although the error bar for this parameter is roughly of the same magnitude as that of $q_0$. Note that there is a slight discrepancy between the mean values for $j_0$ in both models. In contrast, the snap parameter $s_0$ exhibits poor constraint from the FRBs data. While all estimated values for $s_0$ in our analysis are in agreement at $1\sigma$ C.L. between each other and the $\Lambda \rm CDM$ snap $s_0$ value is inside of such interval, it is not possible to discern whether the Universe possesses an evolving dark energy component or a constant one. Consequently, in general, further data and analysis are needed to refine our understanding and provide more precise constraints on the values of $H_0$, $q_0$, $j_0$, and $s_0$.

The contour plots corresponding to the results presented in Table \ref{results1} are displayed in Fig.~\ref{gcont1}.  The findings for both models I and II exhibit striking similarity, with only slight discrepancies observed in the $H_0$-$j_0$, $j_0$-$q_0$, and $j_0$-$s_0$ planes. It is worth noting that, similar to other probes such as supernovae type Ia (SNIa) and cosmic chronometers (CC), the $q_0$-$H_0$ plane derived from FRBs data showcases an anticorrelation trend. Additionally, the $j_0$-$q_0$ plane exhibits some extremely weak correlation features and  interestingly  the $j_0$-$H_0$ plane exhibits a positive correlation different  than anticorrelation in SNIa and CC case. These  patterns suggest that tests employing FRBs data could arise as complementary to other probes at the background level. We will delve further into this point in the subsequent discussion.

\begin{center}
%\begin{table*}
\captionof{table}{The constraints for the cosmographic expansion up to the snap parameter. In this case, we assume a gaussian prior to $\Omega_b h^2$ and a narrow prior over $H_0$.}
\resizebox{\linewidth}{!}{
\begin{tabular}{c c c c c}
\hline%\toprule
Model & $q_0$ & $j_0$ & $s_0$ & $f_{\rm IGM}$ \\
\hline%\midrule
\RomanNumeralCaps{1} & $-0.61\pm0.19$ & $0.67${\raisebox{0.5ex}{\tiny$\substack{+0.37 \\ -0.50}$}} & $-4.6\pm6.6$ & $0.84\pm0.06$\\
\RomanNumeralCaps{2} & $-0.59${\raisebox{0.5ex}{\tiny$\substack{+0.23 \\ -0.16}$}} & $1.02\pm0.56$ & $-1.7\pm6.7$ & $0.82\pm 0.07$\\
\hline
%\bottomrule
\end{tabular}}
%\end{table*}
\label{orderexp}
\end{center}

On the other hand, in case b), we consider a Gaussian prior for $\Omega_b h^2$ with a value of $0.02235 \pm 0.00049$, as reported by \citet{cooke2018one}. Additionally, we impose a narrow flat prior on $H_0$ within the range $\left[66, 75\right]$, in line with the framework addressing the Hubble tension. This choice allows us to estimate the baryon fraction in the intergalactic medium $f_{\rm IGM}$, alongside $q_0$, $j_0$, and $s_0$. The results for this case are summarized in Table \ref{orderexp}. Regarding $f_{\rm IGM}$, both models I and II agree within a $1\sigma$ confidence level, as well as with previous studies. The analysis suggests that approximately $82\%$ of the baryons are accounted for in the late-time Universe, consistent with the findings of \citet{shull2012baryon}. The precision of this result is approximately $7\%$ for model I and $8\%$ for model II. In terms of $q_0$ and $j_0$, the results are consistent in both models. Despite the high uncertainties, all the results remain consistent with each other. Our results are in accordance with those in \citet{shull2012baryon}. Their analysis suggests that approximately $18\pm4\%$ of the baryons in the Universe are in a collapsed form, with $7\pm2\%$ residing in galaxies, $4\pm1.5\%$ in the intercluster medium (ICM), $5\pm3\%$ in the circumgalactic medium (CGM), and $1.7\pm0.4\%$ in cold neutral gas clouds, estimating that roughly $100\%$ of the baryons are found in low redshift.

Considering the limited constraints and theoretical motivations for $\mathrm{DM}_\mathrm{host}$, we also explored scenarios where the mean values of the distributions for this component, denoted as $\langle \mathrm{DM}_\mathrm{host} \rangle$ and $e^{\mu}$ in models $\mathrm{I}$ and $\mathrm{II}$ respectively, are treated as free parameters. Our investigation reveals that incorporating it them as a fixed value, weighted by the redshift of the source, yields comparable results without significant differences. 

\subsection{Combined constraints with CC, SNIa, and FRBs}
%%%%%%%%%%%%%%%%%%%%%%%%%%%%%%%%%%%%%%%

To compare our results with other results in the literature, we use 27 of the 41 measurements of the Hubble parameter H(z) compiled by \citet{jesus2018bayesian}, inferred through the cosmic chronometers technique and also using the position of the peak of Baryon Acoustic Oscillations, which provides a standard rule in the radial direction when measuring the distribution of galaxies by mapping the large scale structure.
The approach presented in \citet{jimenez2002constraining} relies on measuring the age difference $\Delta t$ between pairs of old spiral galaxies that formed at comparable times but are separated by a distance $\Delta z$. This method is particularly suitable for galaxies with low levels of stellar dust, such as the selected spiral galaxies, which allows for easier acquisition of their luminous spectra \citep[]{padilla2021cosmological}. By applying this technique, researchers can estimate the Hubble parameter using the following relationship:
\begin{equation}
    H(z)=-\frac{1}{1+z}\frac{dz}{dt}\simeq-\frac{1}{1+z}\frac{\Delta z}{\Delta t}.
\end{equation}

By observing galaxies at remote times, we can use the age evolution of their stars as a clock to measure cosmic time. This method has a significant advantage as it avoids systematic errors that may arise when measuring the absolute ages of individual galaxies, instead allowing for the measurement of the relative age difference ($dt$) between them. Additionally, this approach enables the independent inference of the Hubble parameter, without relying on a specific cosmological model, as pointed out by \citet{negrelli2020testing}. To estimate the cosmographic parameters from Cosmic Chronometers, we need to evaluate the likelihood function, computed as follows:

\begin{equation}
    \mathcal{L}_{\rm CC} \propto  \exp \left[ \frac{\bigl(H_i - H^{\rm{theo}}(z_i) \bigr)^2} {2 \sigma_{H_i}^2} \right] \,,
\end{equation}

\noindent where $H_i$ represents each of the individual measurements of the sample considered and $H^{\rm theo}$, are the set of Hubble parameters calculated from the cosmographic expansion, see \citet{lizardo2021cosmography}.
%\subsection{Type Ia Supernova}

We also include supernovae type Ia (SNIa) in our analysis, covering the same redshift range as the FRBs in Table 1. Specifically, we construct a subsample of the Pantheon catalogue, which consists of 926 data points. The Pantheon dataset comprises a comprehensive collection of SNIa observations, incorporating data from various surveys such as PanSTARRS1, SDSS, SNLS, and various low-z and HST samples \citep[][]{scolnic2018complete}. The redshift range of the SNIa data spans from $0.01$ to $2.26$. The distance modulus for SNIa is defined as follows:
%Supernovae are highly energetic events that occur in stars at the
%end of their lives, resulting in bright explosions and the production
%of heavy elements essential to life. They are classified based on their
%spectral features and light curves, with Type Ia supernovae (SNIa)
%being particularly important for cosmological research. 
\begin{equation}
    \mu_{\rm obs}=m_{\rm B}-M_{\rm B},
\end{equation}
being $m_{\rm B}$ the apparent magnitude and $M_{\rm B}$ the absolute magnitude. Using the luminosity distance relation,
\begin{equation}
    d_{\rm L}=\left(1+z\right)\frac{c}{H_0}\int_0^z\frac{dz^\prime}{E(z^\prime)},
\end{equation}
the theoretical distance modulus is expressed as:
\begin{equation}
    \mu_{\rm th}=5\log_{10}\left(\frac{d_{\rm L}}{\rm Mpc}\right)+25.
\end{equation}
By fitting this quantity, it is possible to obtain constraints from SNIa through the relation:

\begin{equation}
     \mathcal{L}_{\rm SNIa} \propto  \exp \left[-\frac{1}{2}(\mathbf{\mu}_{\text{obs}} - \mathbf{\mu}_{\text{th}})^T \mathbf{C}^{-1} (\mathbf{\mu}_{\text{obs}} - \mathbf{\mu}_{\text{th}})\right],
\end{equation}

\noindent where $\mu_{obs}$ are the observed distance modulus and $C$ is the covariance matrix of the events, check \citet{scolnic2018complete} for more information.

\begin{figure*}
 \includegraphics[width=15cm]{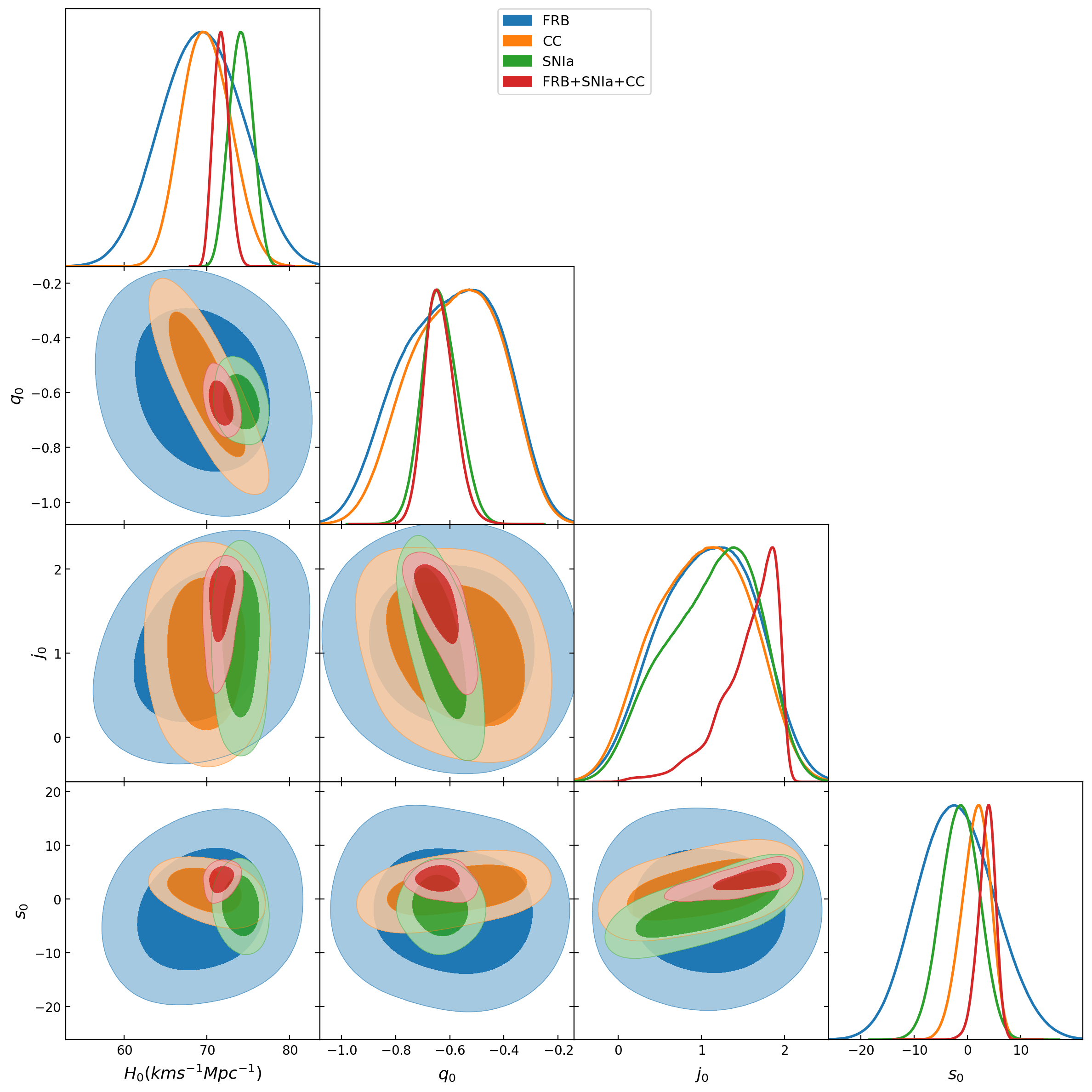}
 \caption{The posterior distribution with the $1\sigma$ and $2\sigma$ statistical confidence regions for the parameters $H_0,~q_0,~j_0,$ and $s_0$, constrained by FRBs, CC, SNIa.}
 \label{gcont}
\end{figure*}
 
Considering that model II has more theoretical motivations and simulations support it, we have chosen this model for the joint analysis. Nevertheless, it is important to note that even working with model II, the limited number of FRB data and the striking similarity between models I and II results (see Table \ref{2}), suggest that there is no huge difference in working with any of them. To perform the joint analysis, we combine the FRB dataset with Cosmic Chronometers and type Ia Supernovae. In this case, the free parameters are $\theta=\{H_0, \Omega_{\rm b} h^2, q_0, j_0, s_0, M_B\}$, with a Gaussian prior over $M_B$, with a mean value of $-19.24$, and a standard deviation of $0.04$ given by \citet[][]{camarena2021use}. The total likelihood then is the product:

\begin{equation}
    \mathcal{L}_{\rm TOT} = \mathcal{L}_{\rm FRB}\times \mathcal{L}_{\rm CC}\times \mathcal{L}_{\rm SNIa}.
\end{equation}

The statistical results are summarized in Table \ref{tab:continued} and displayed in Fig. \ref{gcont}. Our combined constraint for Hubble constant $H_0$ achieves a precision of $\sim 1\%$. It is worth noting a preference for lower values of $H_0$  when utilising Cosmic Chronometers. This preference holds whether employing the cosmography technique or working within the framework of the $\Lambda\rm CDM$ model \citep[][]{moresco2022unveiling, busti2014evidence} and, as a result, a subtle tension of  $ \sim 1.2\sigma$ arises in comparison to the constraints obtained from type Ia Supernovae, causing such a small precision
in the joint constraint. The deceleration parameter $q_0$ achieves a precision of  $\sim 9\%$ while the jerk $j_0$ has a precision of $\sim 26 \%$. Furthermore, it is worth mentioning that although the snap parameter $s_0$ does not align within the $1\sigma$ region of the $\Lambda \rm CDM$ model, the inclusion of FRBs in conjunction with Cosmic Chronometers and type Ia Supernovae reduces the error associated with this parameter by approximately $30\%$.

%we highlight that the credible regions for $j_0\times H_0$, in figure \ref{gcont}, suggests that the FRB can be a complementary test to the others (SNIa and CC), so the combination of them should be a great probe for these parameters in the future.

%Our estimated value for the jerk parameter is $1\sigma$ consistent with the fiducial value. \mvs{fiducial value? é uma simulação por acaso?}%For instance, the tension between the values for $q_0^\prime$ calculated using model $\rm \RomanNumeralCaps{1}$ and model $\rm \RomanNumeralCaps{2}$ for the gaussian prior is less than $1\sigma$.

\begin{table}
\begin{center}
 \caption{Constraints for all the parameters using the model I with a gaussian prior on $\Omega_b h^2$ and fixing $f_{\rm IGM} = 0.82$.}
 \label{tab:continued}
 \resizebox{\linewidth}{!}{
\begin{tabular}{c c c c c}
\hline
Parameter & FRB & CC & SNIa & FRB+CC+SNIa\\
\hline
$H_0$ & $69.4\pm 4.7$ & $69.9\pm 2.9$ & $74.1\pm1.4$ & $71.6${\raisebox{0.5ex}{\tiny$\substack{+0.9 \\ -1.0}$}}\\
$q_0$ & $-0.59${\raisebox{0.5ex}{\tiny$\substack{+0.20 \\ -0.17}$}} & $-0.57\pm 0.16$ & $-0.64${\raisebox{0.5ex}{\tiny$\substack{+0.05 \\ -0.07}$}} & $-0.64${\raisebox{0.5ex}{\tiny$\substack{+0.05 \\ -0.06}$}}\\
$j_0$ & $1.08${\raisebox{0.5ex}{\tiny$\substack{+0.62 \\ -0.56}$}} & $1.01\pm0.55$ & $1.13${\raisebox{0.5ex}{\tiny$\substack{+0.66 \\ -0.47}$}} & $1.57${\raisebox{0.5ex}{\tiny$\substack{+0.42 \\ -0.15}$}}\\
$s_0$ & $-2.1\pm7.0$ & $1.6${\raisebox{0.5ex}{\tiny$\substack{+2.9 \\ -2.4}$}} & $-1.3\pm3.5$ & $3.6${\raisebox{0.5ex}{\tiny$\substack{+1.8 \\ -1.4}$}} \\
\hline
\end{tabular}}
\end{center}
\end{table}

%\begin{table}
%	\centering
%	\caption{These are the values estimated by using a gaussian prior on $\Omega_b h^2$.}
%	\label{obgauss}
%	\begin{tabular}{cccc} % four columns, alignment for each
%		\hline
%		Model & $H_0$ & $q_0^\prime$ & $\Omega_b h^2$\\
%		\hline
%		\RomanNumeralCaps{1} & $68.7\pm 5.9$ & $-0.42 \pm 0.18$ & $0.022348 \pm 0.000473$\\
%		\RomanNumeralCaps{2} & $69.4 \pm\ 7.8$ & $-0.18 \pm 0.31$ & $0.022340 \pm 0.001012$\\
%		\hline
%	\end{tabular}
%\end{table}

%\begin{figure}
%    %\subfloat[\centering ]
%    \includegraphics[width=\columnwidth]%{plotgaussprior.png}
%    \label{gaussiancont}
%\end{figure}

%\begin{table}
%%	\centering
%	\caption{These are the values estimated by using a flat prior on $\Omega_b h^2$.}
%	\label{tab:example_table}
%	\begin{tabular}{cccc} % four columns, alignment for each
%		\hline
%		Model & $H_0$ & $q_0^\prime$ & $\Omega_b h^2$\\
%		\hline
%		\RomanNumeralCaps{1} & $68.0\pm 5.8$ & $-0.42${\raisebox{0.5ex}{\tiny$\substack{+0.13 \\ -0.25}$}} & $0.02236 \pm 0.00028$\\
%		\RomanNumeralCaps{2} & $69.4${\raisebox{0.5ex}{\tiny$\substack{+6.4 \\ -8.2}$}} & $-0.21 \pm 0.29$ & $0.02235 \pm 0.00029$\\
%		\hline
%	\end{tabular}
%\end{table}

%%%%%%%%%%%%%%%%%%%%%%%%%%%%%%%%%%%%55555

%%%%%%%%%%%%%%%%%%%%%%%%%%%%%%%%%%%%%%5
\section{Discussion}\label{6}
%%%%%%%%%%%%%%%%%%%%%%%%%%%%%%%%%%%%%%%5
Fast Radio Bursts (FRBs) have become an increasingly valuable tool in cosmological research, with the $\rm DM - z$ relation being employed in numerous previous studies. In this work, we expand upon this analysis by introducing the cosmographic expansion of the $\rm DM - z$ expression. To assess the reliability of our approach, we consider two distinct models. The first model incorporates straightforward assumptions and treats the host and intergalactic medium (IGM) components as Gaussian distributions. The second model is more intricate, assuming a quasi-Gaussian distribution for the IGM component and a lognormal distribution for the host component. Our results indicate that both approaches yield significant improvements over previous studies that solely relied on the $\rm DM - z$ relation, achieving an interesting precision of $6\%$ for model I and $7\%$ for model II for the Hubble parameter ($H_0$) despite the limited number of measured FRBs. According to the standard error formula, we estimate that approximately 2400 FRBs with redshift measurements are required to achieve a similar precision level as the SH0ES collaboration. Fortunately, this goal is within reach in the near future, given the growing number of instruments dedicated to the detection of FRBs. Two prominent upcoming radio telescopes deserve mention: the Square Kilometre Array (SKA) and the Baryon Acoustic Oscillations from Integrated Neutral Gas Observations (BINGO) project. The SKA, renowned for its sub-arcsecond accuracy \citep[][]{zhang2023cosmology}, promises to revolutionize FRB (Fast Radio Burst) research by enabling the detection of thousands of FRBs and their redshift counterparts \citep[][]{macquart2015fast}. On the other hand, the BINGO project, currently under construction in Brazil \citep[][]{abdalla2022bingo, santos2023bingo}, focuses specifically on detecting the 21-cm line of neutral hydrogen (HI) and is also expected to be a formidable instrument for FRB detection.

By employing the cosmographic expansion, we successfully estimated the kinematic parameters $q_0$, $j_0$, and $s_0$ for both models. Notably, our findings exhibited strong agreement between the two models I and II, but it must be recalled that model II has more theoretical foundations and support from cosmological simulations. Utilising only FRBs, we obtained compelling evidence for the acceleration of the expanding Universe. Moreover, the results from both models indicated the presence of a transitional phase during which the dynamics of the universe underwent a change. However, given the limited constraints on the snap parameter, further data are required to discern whether the accelerated expansion is governed by a constant dark energy component or a time-evolving one. 

In addition, the present study has investigated how the cosmographic expansion up to the snap parameter can impact on the estimation of the fraction of baryons in the intergalactic medium (IGM). Our analysis revealed that all the models we considered yielded very consistent results that were in agreement with previous studies. Despite the large statistical errors, we were able to address the missing baryon problem by utilising well-localised FRBs. Our calculations yielded an estimated fraction of baryons in the IGM of $f_{\rm IGM}=0.82\pm0.06$, which suggests that the majority of baryons are indeed accounted for in the IGM. This finding is significant because it provides further insight into the distribution of matter in the universe in a model-independent way and  could have implications for our understanding of cosmic structure formation. 

Based on the results of this study, it is clear that cosmography has played a crucial role in advancing our understanding of FRBs and their usefulness as a cosmological tool. By incorporating higher-order kinematic parameters, such as the snap parameter, we were able to improve the precision of our estimates for the Hubble parameter and the fraction of baryons in the IGM. The high level of consistency between our models and previous studies underscores the reliability of cosmography as a technique for understanding the properties of the universe. Additionally, the continued search for new FRBs and the development of more advanced observational techniques could lead to even more precise estimates of cosmological parameters, further refining our understanding of the evolution of the Universe and its fundamental properties.

\section*{Acknowledgements}

JASF thanks FAPES for financial support. WSHR thanks FAPES (PRONEM No 503/2020) for the financial support under which this work was carried out. The authors thank V. Marra for his insightful comments on this paper.

%%%%%%%%%%%%%%%%%%%%%%%%%%%%%%%%%%%%%%%%%%%%%%%%%%
\section*{Data Availability}

The data used in this work is publicly available at \url{https://www.wis-tns.org/} and \url{http://frbhosts.org/#explore}.

%%%%%%%%%%%%%%%%%%%% REFERENCES %%%%%%%%%%%%%%%%%%

% The best way to enter references is to use BibTeX:

\bibliographystyle{mnras}
\bibliography{example} % if your bibtex file is called example.bib

% Alternatively you could enter them by hand, like this:
% This method is tedious and prone to error if you have lots of references
%\begin{thebibliography}{99}
%\bibitem[\protect\citeauthoryear{Author}{2012}]{Author2012}
%Author A.~N., 2013, Journal of Improbable Astronomy, 1, 1
%\bibitem[\protect\citeauthoryear{Others}{2013}]{Others2013}
%Others S., 2012, Journal of Interesting Stuff, 17, 198
%\end{thebibliography}

%%%%%%%%%%%%%%%%%%%%%%%%%%%%%%%%%%%%%%%%%%%%%%%%%%

%%%%%%%%%%%%%%%%% APPENDICES %%%%%%%%%%%%%%%%%%%%%

%\appendix

%\section{Some extra material}

%

%%%%%%%%%%%%%%%%%%%%%%%%%%%%%%%%%%%%%%%%%%%%%%%%%%

% Don't change these lines
\bsp	% typesetting comment
\label{lastpage}
\end{document}